\documentclass[a4paper]{jpconf}
\usepackage{graphicx}
\usepackage{color}
\usepackage{amssymb}
\usepackage{amsmath}
\usepackage{multirow}
\usepackage{amscd}
\usepackage{verbatim}
\usepackage{pstricks,pst-node,pst-text,pst-3d}
\usepackage{curves}
\usepackage{shadow}

\definecolor{cream}{rgb}{.97, .95, .88}
\definecolor{darkcream}{rgb}{1., .88, .5}
\definecolor{lightpink}{rgb}{0.98, 0.88, 0.87}
\definecolor{lightwhite}{rgb}{1., 0.98, 0.95}
\definecolor{lightsalmon}{rgb}{1., 0.95, 0.90}
\definecolor{lightviolet}{rgb}{0.9, 0.8, 0.9}
\definecolor{lightgray}{rgb}{.96, .96, .96}  
\definecolor{lgray}{rgb}{.75, .75, .75}
\definecolor{LemonChiffon}{rgb}{0.95, 1., 0.7}
\definecolor{lightolivegreen}{rgb}{0.84, 0.89, 0.25}
\definecolor{lightgreen}{rgb}{.664, 1., .52}
\definecolor{llgreen}{rgb}{.900, .983, .960}
\definecolor{tristle}{rgb}{0.87, 0.67, 0.77} 
\definecolor{pink}{rgb}{0.95, 0.45, 0.75}
\definecolor{magenta}{rgb}{1., 0, 1.}
\definecolor{violet}{rgb}{0.9, 0.20, 0.85}
\definecolor{darkolivegreen}{rgb}{0.55, 0.65, 0.35}
\definecolor{maroon}{rgb}{0.7, 0.26, 0.56}
\definecolor{lightmaroon}{rgb}{0.85, 0.38, 0.58}
\definecolor{darkmaroon}{rgb}{0.604, 0.169, 0.451}
\definecolor{ddarkmaroon}{rgb}{0.2, 0.03125, 0.150}
\definecolor{mediumorchid}{rgb}{0.8, 0.33, 0.83}
\definecolor{mediumorchidd}{rgb}{1., 0.33, 0.63}
\definecolor{darkgreen}{rgb}{0.1, 0.6, 0.13}
\definecolor{lightyellow}{rgb}{1., 1., 0.82}
\definecolor{turquoise}{rgb}{0.042, 0.586, 0.512}
\definecolor{turquoisel}{rgb}{0.66, 0.94, 0.83}
\definecolor{darkturquoise}{rgb}{0.21, 0.55, 0.50}
\definecolor{coral}{rgb}{1., 0.6, 0.21}
\definecolor{lightorange}{rgb}{1., 0.88, 0.75}
\definecolor{orangered}{rgb}{1., 0.5, 0.}
\definecolor{orange}{rgb}{1., 0.65, 0.1}
\definecolor{orangel}{rgb}{1., .85, .3}
\definecolor{darkorange}{rgb}{0.875, 0.4, 0.204}
\definecolor{ddarkorange}{rgb}{.675, .218, .05}
\definecolor{bluesky}{rgb}{0.48, 0.53, 1.}
\definecolor{gold}{rgb}{1., 0.85, 0.25}
\definecolor{goldd}{rgb}{0.95, 0.75, 0.05}
\definecolor{darkviolet}{rgb}{0.54, 0.04, 0.84}
\definecolor{ddarkviolet}{rgb}{.382, .063, .657}
\definecolor{lightblue}{rgb}{0.30, 0.86, 0.89}
\definecolor{LightBlue}{rgb}{0.68, 0.85, 0.9}
\definecolor{lblue}{rgb}{0.78, 0.90, 0.95}
\definecolor{darkblue}{rgb}{.105, .308, .707}
\definecolor{lightmaroon}{rgb}{0.85, 0.38, 0.58}
\definecolor{darkmaroon}{rgb}{0.604, 0.169, 0.451}
\definecolor{darkpink}{rgb}{0.879, 0.020, 0.766}
\definecolor{ddarkpink}{rgb}{0.738, 0.195, 0.406}
\definecolor{grey}{rgb}{0.717, 0.717, 0.717}
\definecolor{lightgrey}{rgb}{0.800, 0.800, 0.800}
\definecolor{brown}{rgb}{0.740, 0.323, 0.182}
\definecolor{redbrown}{rgb}{.575, .158, .05}
\definecolor{darkbrown}{rgb}{0.34, 0.25, 0.05}
\definecolor{orangebrown}{rgb}{0.433, 0.262, 0.06}
\definecolor{pinkl}{rgb}{1., 0.788, 0.918}
\definecolor{salmon}{rgb}{1., 0.66, 0.5}
\definecolor{lightbrown}{rgb}{0.703, 0.508, 0.121}

\def\Journal#1#2#3#4{#3 {#1} {\bf #2} #4}

\def\JHE{\em J. High Ener. Phys.}

\def\MPA{{\em Mod. Phys.} A}
\def\MPL{{\em Mod. Phys. Lett.} A}

\def\NPA{{\em Nucl. Phys.} A}

\def\PRD{{\em Phys. Rev.} D}

\def\PRV{\em Phys. Rev.}

\def\PTPS{\em Prog. Theor. Phys. Suppl.}

\def\RMP{\em Rev. Mod. Phys.}
\def\RPP{\em Rept. Prog. Phys.}

\def\be{\begin{equation}}
\def\ee{\end{equation}}
\def\bea{\begin{eqnarray}}
\def\eea{\end{eqnarray}}
\def\bes{\begin{equation*}}
\def\ees{\end{equation*}}
\def\beas{\begin{eqnarray*}}
\def\eeas{\end{eqnarray*}}

\def\mg{\mathsf g}

\begin{document}
\title{Quantum coherent states in cosmology}
\author{Houri~Ziaeepour}
\address{Institut UTINAM-CNRS, Universit\'e de Franche Comt\'e, Besan\c{c}on, France}
\ead{houriziaeepour@gmail.com}

\begin{abstract}
Coherent states consist of superposition of infinite number of particles and do not have a 
classical analogue. We study their evolution in a FLRW cosmology and show that only when full 
quantum corrections are considered, they may survive the expansion of the Universe and form a 
global condensate. This state of matter can be the origin of accelerating expansion of the 
Universe, generally called dark energy, and inflation in the early universe. Additionally, 
such a quantum pool may be the ultimate environment for decoherence at shorter distances. 
If dark energy is a quantum coherent state, its dominant contribution to the total energy of the 
Universe at present provides a low entropy state which may be necessary as an 
initial condition for a new Big Bang in the framework of bouncing cosmology models.
\end{abstract}

\section{Introduction}
Cosmological observations have demonstrated that at least during two epochs the Universe has had an 
accelerating expansion. The first era is called {\it inflation}~\cite{infrev} and occurred after 
the Big Bang. The second era is the present accelerating expansion which its effect has become 
significant since redshift $~0.5$. Its origin is unknown and is given the generic name of 
{\it dark energy}~\cite{derev}. It is generally believed that inflation is caused by quantum 
phenomena such as symmetry breaking and phase transition, most probably related to the breaking 
of supersymmetry or supergravity. The smallness of limits put on primordial non-gaussianity by 
observations~\cite{cmbplancknongauss} is probably an evidence for a high energy scale of inflation 
$E_{inf} \sim 10^{-3}- 10^{-6} M_P$ where $M_P$ is the Planck mass~\cite{infscale}. 

Dark energy may be explained in the same manner, that is induced by condensation of a quantum 
scalar field - or a vector field in some models. Although in the last few years or so gravity 
models have become a favorite candidate for the origin of dark energy, the fact that inflation 
and recent acceleration of the Universe seem to be very similar encourage the hypothesis of a 
quantum origin rather than geometry for dark energy. Obviously the underlying quantum model may 
be related to quantum gravity - if there is at all such a model. But it is not a necessarily so. 
In fact it has been shown that condensation of quantum fields at cosmological scales can occur 
after inflation and reheating, that is during radiation and matter domination 
epochs~\cite {houridmquin,hourifat,houricond}. Moreover, a non-gravitational interaction in the 
dark sector is a physically motivated possibility. For instance the decay of a heavy dark matter 
which is produced during preheating and has a small branching ratio to a very light scalar field, 
provides a {\it quintessence field} to play the role of dark energy. Moreover, it explains in a 
natural way and without violation of null energy principle the effective equation of state of 
dark energy $w_{ed}$ which seems to be $\lesssim -1$~\cite{houridmdecay}.

In this proceedings we first review condensates and their relation with coherent states. The 
latter can be used to introduce a frame independent definition for vacuum of quantum fields in 
curved spacetimes. We show that according to this definition the energy of the real vacuum is always 
zero. This solves the apparent problem of too small observed vacuum if dark energy is interpreted 
as the energy of the vacuum. Assuming that dark energy is the condensate of one or more quantum 
fields, we briefly review its evolution in an expending Universe and discuss the necessary 
conditions for its survival. Finally we present preliminary results from an on going study which 
its aim is to investigate the evolution of cosmological condensates in the frame work of 
non-equilibrium quantum field theory.

\section {Quantum condensates and coherent states}
\subsection {Definition of quantum condensates}
Condensates of quantum scalar fields are very special states of matter. They are defined as states 
$|\psi\rangle$ for which:\footnote{Here we present fields and states in interaction picture unless 
explicitly specified.}
\be
\langle \psi |\Phi| \psi \rangle \neq 0 \label {conddef}
\ee
By decomposing the field $\Phi$, which evolves according to free Hamiltonian of the theory, to 
annihilation and creation operators, it is straightforward to see that the state $|\psi \rangle$ 
cannot contain a finite number of particles. Condensate states are quantum field theoretical 
extension of Bose-Einstein Condensate (BEC) in condensed matter when the number of particles 
sharing the same energy state in very large. The list of the best observed examples of such states 
include: the condensate of Cooper pair of electrons in superconductors, condensate of 
electron-hole called exciton in semiconductors, quark-antiquark condensate in hadronic matter 
leading to the breaking of chiral symmetry, and condensation of Higgs field leading to the breaking 
of Electroweak symmetry and mass acquisition Standard Model (SM) particles. Finally, the most 
recent observations of Cosmic Microwave Background (CMB) is consistent with an spinorial 
condensate, that is a condensate with effective negative mass and two minima in its effective 
potential~\cite{infcmbplackpoten}. There is no general expression for a condensate state. 
Nonetheless, coherent states fulfill the definition (\ref{conddef}).

\subsection{Coherent States}
R.J. Glauber~\cite {coherestate} has proposed 3 equivalent definitions for condensate 
states~\cite{coherestaterev}:
\begin{enumerate}
\item Eigen state of annihilation operator $a_\alpha$:
\be
a_\alpha |\psi \rangle = C_\alpha |\psi \rangle, \quad [a_\alpha,a_\alpha'^\dagger] = 
\delta_{\alpha\alpha'} \label{coher1}
\ee
\item State generated by application of displacement operator $D_\alpha (C_\alpha)$ on the vacuum 
(a reference state):
\be
|C_\alpha \rangle = D_\alpha (C_\alpha) |0 \rangle, \quad 
D_\alpha (C_\alpha) \equiv \exp (C_\alpha a_\alpha - C^*_\alpha a^\dagger_\alpha) \label{coher2}
\ee
\item State with minimum uncertainty: 
\bea
&& (\Delta q)^2 (\Delta p)^2 = (\hbar/2)^2, \quad \hat{q} \equiv (a + a^\dagger) / 
\sqrt {2}, \quad \hat{p} \equiv (a - a^\dagger) / i\sqrt{2}, \nonumber \\
&& (\Delta f)^2 \equiv \langle \psi|(\hat{f} - \langle \hat{f} \rangle)^2| \psi \rangle, \quad 
\langle \hat{f} \rangle \equiv \langle \psi |\hat{f}| \psi \rangle \quad \text {for any operator
$\hat{f}$} \label{coher3}
\eea
\end{enumerate}
where $\alpha$ indicates indices that classify different modes, including symmetry indices in 
multi-field models. Coherent states that satisfy (\ref{coher1}) fulfill also the condition for a 
condensate state. In particular, Glauber coherent state~\cite{coherestate} is defined as: 
\be
|C_\alpha \rangle \equiv e^{-|C_\alpha|^2} e^{C_\alpha a^\dagger_\alpha} = e^{-|C_\alpha|^2} \sum_{i=0} C^i_\alpha / 
i! (a^\dagger_\alpha)^i
\ee
and satisfies all the definitions (\ref{coher1}-\ref{coher3}) as well as the condensate 
condition (\ref{conddef}). On the other hand, various extensions of this state may satisfy only 
some of the definitions~\cite{coherestaterev}. The generalization that we will review in the next 
subsection is based on (\ref{coher1}) because it explicitly satisfies the definition of a 
condensate state.

\subsection {\bf A generalized coherent state}
In~\cite{houricond} we described a generalized coherent state for a scalar field based on 
Glauber coherent state:
\bea
&& |\Psi_{GC}\rangle \equiv \sum_k A_k e^{C_k a_k^{\dagger}} |0\rangle = \sum_k A_k 
\sum_{i=0}^{N \rightarrow \infty} \frac {C_k^i}{i!}(a_k^{\dagger})^i |0\rangle \label{condwaveg} \\ 
&& a_k |\Psi_{GC}\rangle = C_k |\Psi_{GC}\rangle \quad \quad 
\langle \Psi_{GC}| N_k |\Psi_{GC}\rangle = |A_k C_k|^2 \label{condwavegann}
\eea
In (\ref{condwaveg}) we have restricted mode indices to momentum $k$. In this case the condensate 
in (\ref{condwaveg}) can be considered as superposition of Glauber condensates moving with 
respect to mode $k = 0$ with a momentum equal to $k$. 

One can further extend the definition (\ref{condwaveg}) by considering products of states of type 
$|\Psi_{GC}\rangle$ which may in addition belong to different field species indicated by index 
$i$ in the following expression: 
\be
|\Psi_{G}\rangle \equiv \sum_{k_1,k_2,\cdots} \biggl (\prod_{k_i} A_{k_i} \biggr) e^{\sum_i C_{k_i} a_{k_i}^{\dagger}} 
|0\rangle \label{condwave}
\ee
Such a state includes products of states in which particles do not have the same momentum and 
consists of all combinations of states with any number of particles and momenta. Because 
(\ref{condwave}) includes linear superposition and direct product of Glauber states, it satisfies 
at least the first two definitions of a coherent state. It is less straightforward to prove the 
third property.

\section{Definition of vacuum using coherent states}
The aim for definition of an extended coherent state~\cite{houricond,hourivacuum} is finding a 
frame independent vacuum state for applications in curved spacetimes. In quantum field theory 
vacuum is defined as particle-less state of a system: 
\be
\hat {N}_\alpha|0\rangle = 0~~\forall~\{\alpha\} \label{vacdef}
\ee
In Minkowski space this state is both physically and mathematically well defined. However, as it 
is well known in curved spacetimes $|0\rangle$ is not preserved under a general coordinate 
transformation and it is projected to a state with infinite number of particles moving with a 
variable speed with respect to each others, see e.g.~\cite{qftcurvebook} and references therein. 
Another definition of vacuum used in the literature is the lowest available energy state. This 
definition has many practical applications. For instance, in condensed matter the lowest energy 
level of conducting Cooper pairs may be called the {\it vacuum}. But is only a relative vacuum 
and does not mean the absence of matter. On the other hand, in cosmology and in the frame work of 
Einstein gravity, every matter component is observable through its gravity and contributes to the 
evolution of spacetime. Therefore, a true vacuum state must be {\it a state with no contribution 
in semi-classical Einstein equation}. Moreover, such a state must be at least asymptotically 
measurable by all observers. Such a state can be used as an absolute {\it reference} for 
verification of the presence of {\it matter}. Evidently, the standard definition of vacuum state 
does not fulfill these conditions because it is frame dependent.

For defining a frame independent vacuum we use the general condensate state (\ref{condwave}).
When $C_{k_i} \rightarrow 0~\forall~k_i$, this state is neutralized by all annihilation operators and 
the expectation value of particle number approaches zero for all modes. Therefore, this state 
satisfies the condition (\ref{vacdef}) for a vacuum state. Coefficients $A_k$ are relative 
amplitude of modes $k$ and are determined by underlying model. Therefore, in general they are not 
zero. The coherent state (\ref{condwave}) has the interesting property that under a Bogolubov 
transformation it is projected to itself:
\be
a_{k_i} = \sum_j \sum_{k_j} {\mathcal A}_{k_j k_i} a'_{k_j} + \sum_j \sum_{k_j} 
{\mathcal B}_{k_j k_i} {a'}^\dagger_{k_j} \quad \quad a^\dagger_{k_i} = \sum_j \sum_{k_j} 
{\mathcal A}^*_{k_j k_i} {a'}^\dagger_{k_j} + \sum_j \sum_{k_j} {\mathcal B}^*_{k_j k_i} a'_{k_j}
\label {bogoltrans}
\ee
Replacing $a_{k_i}^{\dagger}$ in (\ref{condwave}) with the corresponding expression in 
(\ref{bogoltrans}) leads to an expression for $|\Psi_{G}\rangle$ similar to (\ref{condwave}) but 
with respect to the new operator ${a'}^\dagger_k$ and $C'_{k_j} = \sum_i \sum_{k_i} 
{\mathcal A}^*_{k_j k_i} C_{k_i}$. For $C_{k_i} \rightarrow 0~\forall~k_i$ and finite 
${\mathcal A}^*_{k_j k_i}$, $C'_{k_i} \rightarrow 0~\forall~k_i$. Note that here we assume that the 
Bogolubov transformation changes $|0\rangle$ to a similar state which is neutralized by 
$a'_k~\forall~k$. Therefore, in contrast to the null state of the Fock space, $|\Psi_{G}\rangle$ 
is frame-independent. However, it is easy to verify that this new definition of vacuum does not 
solve the singularity of naive determination of the expectation value of energy-momentum tensor 
$\hat{T}^{\mu\nu}$ without operator ordering~\cite{weinbergde, hourivacuum}. The singularity of 
$\hat{T}^{\mu\nu}$ originates from extension of classical $T^{\mu\nu}$ to a quantum field theory 
operator functional. Due to crossing ambiguity, this operator is not well defined and needs a 
regularization or operator ordering, but this operation is considered to be not appropriate in 
the context of Einstein model of gravity in which there is an absolute energy reference and all 
non-zero energies affect the spacetime metric. 

Rather than using $\hat{T}^{\mu\nu}$, it is proposed to use number operator $\sum_k \hat{N}_k$ to 
determine the energy density of quantum states and vacuum~\cite{qftcurvebook}. Notably, 
frame-independent of vacuum defined above using general coherent state (\ref{condwave}) 
is neutralized by the number operator $\hat {N}_k |\Psi_{G}\rangle = 0 ~\forall~k$. Previously, 
this alternative method for measuring energy density of vacuum has been considered to be a poor 
replacement because the standard definition of vacuum in (\ref{vacdef}) is frame dependent. In 
addition to an ambiguity-free determination of vacuum energy, the new definition is more consistent 
with many-particle nature of quantum field theory and can be experimentally realized as a condensate 
with an amplitude approaching zero. 

\section{Dark energy and condensates}
The absolute and frame independent vacuum defined above has a null energy according to the 
prescription explained in the previous section. Therefore, the observed accelerating expansion 
of the Universe cannot be due to such a vacuum and we must find alternative explanations. Many of 
suggested alternatives are based on classical scalar fields. Considering the fact that we live 
in a quantum universe, ultimately the origin of the classical field is quantum. Even in modified 
gravity models in which the classical scalar field is associated to geometry, presumable dilaton 
in a scalar-tensor gravity models, the origin of deviation from Einstein gravity is assumed to 
be quantum gravity, e.g. supergravity, models inspired by string theory, etc.

The only way to obtain a classical field $\varphi (x)$ from a quantum one is its condensation, 
that is a state for which $\langle \psi | \Phi (x) | \psi \rangle \equiv \varphi (x) \neq 0$. 
As we discussed in the previous sections $|\psi\rangle$ is a condensate and we must study the 
evolution of a quantum scalar field in a FLRW geometry to see whether a condensate with properties 
of dark energy can form. Many authors have studied the evolution of quantum perturbations of scalar 
field(s) and particle production in the framework of quantum field theory in de Sitter spacetime 
as a good approximation for geometry of the Universe during inflation~\cite{infscalar}. The 
evolution of condensate(s) is also studied for some inflation model~\cite{infcond}, but its 
formation and evolution pre-inflation era and backreaction of its evolution and particle production 
on metric is usually ignored.

The study of the evolution of dark energy candidate scalar fields is more sophisticated because it 
must cover more recent epochs of the Universe, namely radiation and matter domination eras, and 
present epoch in which dark energy is dominant but matter also has yet a non-negligible 
contribution. In these eras the geometry is not as symmetric as de Sitter and thereby it is more 
difficult to solve dynamical equations. Moreover, to have a consistent solution which can be 
compared with observation one needs to consistently determine the backreaction of various 
components on each others.

We have begun the programme of studying the condensation of a quintessence field\footnote {For 
the sake of simplicity, here we call all dark energy candidate scale fields {\it quintessence}, 
even when the field belong to a modified gravity model, as long as it is considered to be a quantum 
field.} with~\cite{houricond} in which we studied the process of condensation in radiation and 
matter domination eras using formulation and techniques of non-equilibrium quantum field theory. 
In this work we limited ourselves to lowest order of quantum corrections, i.e. tree diagrams. A full 
formulation using 2-Particle-Irreducible (2PI) technique and its numerical simulation is at present 
an on going work. Here we briefly review conclusions of analytical investigations and preliminary 
results from numerical simulations.

\subsection{Formation of a condensate}
We study, in the framework of 2PI formalism, the evolution of a model with 3 scalars each in one 
of 3 physically important scales of particle physics: {\bf A decaying massive particle $X$:} 
Depending on the initial condition it may be 
considered as inflaton or a decaying dark matter, for initial condition fixed at pre-inflation 
epoch or post inflation and preheating, respectively. In both cases the mass scale of the 
particle is assumed to be (sub)GUT (Grand Unification Theory) scale; 
{\bf A scalar field with electroweak scale mass $A$:} This is assumed to be one of the 
two remnants of the decay of $X$ particles, presumably a Standard Model particle or one which 
ultimately decay to SM particles. It can be also a collective notation for all remnants except a 
light axion-like scalar field; {\bf A light axion-like (pseudo-)scalar $\Phi$:} The condensate 
of this field should play the role of a classical quintessence field. 

We assume that heavy fields $X$ and $A$ do not have self-interaction and produce negligible 
condensate. Indeed in~\cite{houricond} it is shown that the amplitude of condensate decays very 
quickly with increasing mass. Because a condensate includes states with infinite number of 
particles, the amplitude of condensate must decreases very quickly with mass, otherwise even a 
very small probability for highly occupied states may make the Universe over-dense.

\begin{figure}
{\color{darkblue}
\begin{tabular}{p{2cm}p{1.5cm}p{2cm}p{2cm}p{0.5cm}p{2cm}p{1cm}}
$\Gamma_2 = \sum\limits_{i=2}^n N_1$ & \includegraphics[width=2cm]{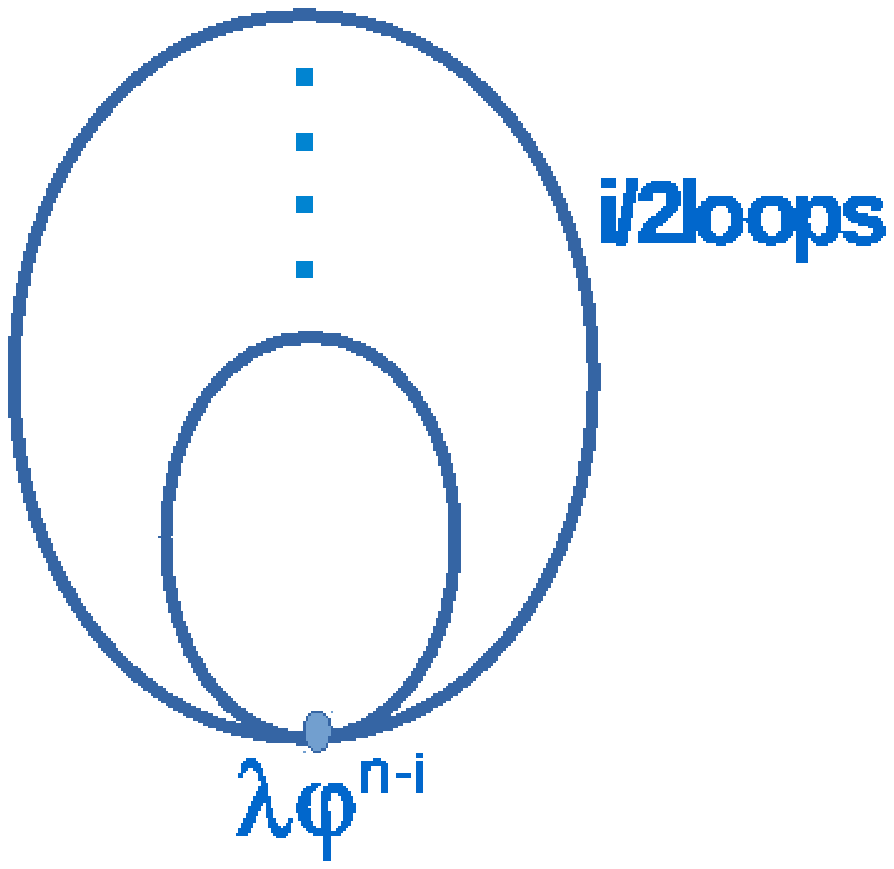} &  
 $+ \sum\limits_{i=2}^n N_2$ & \includegraphics[width=2.5cm]{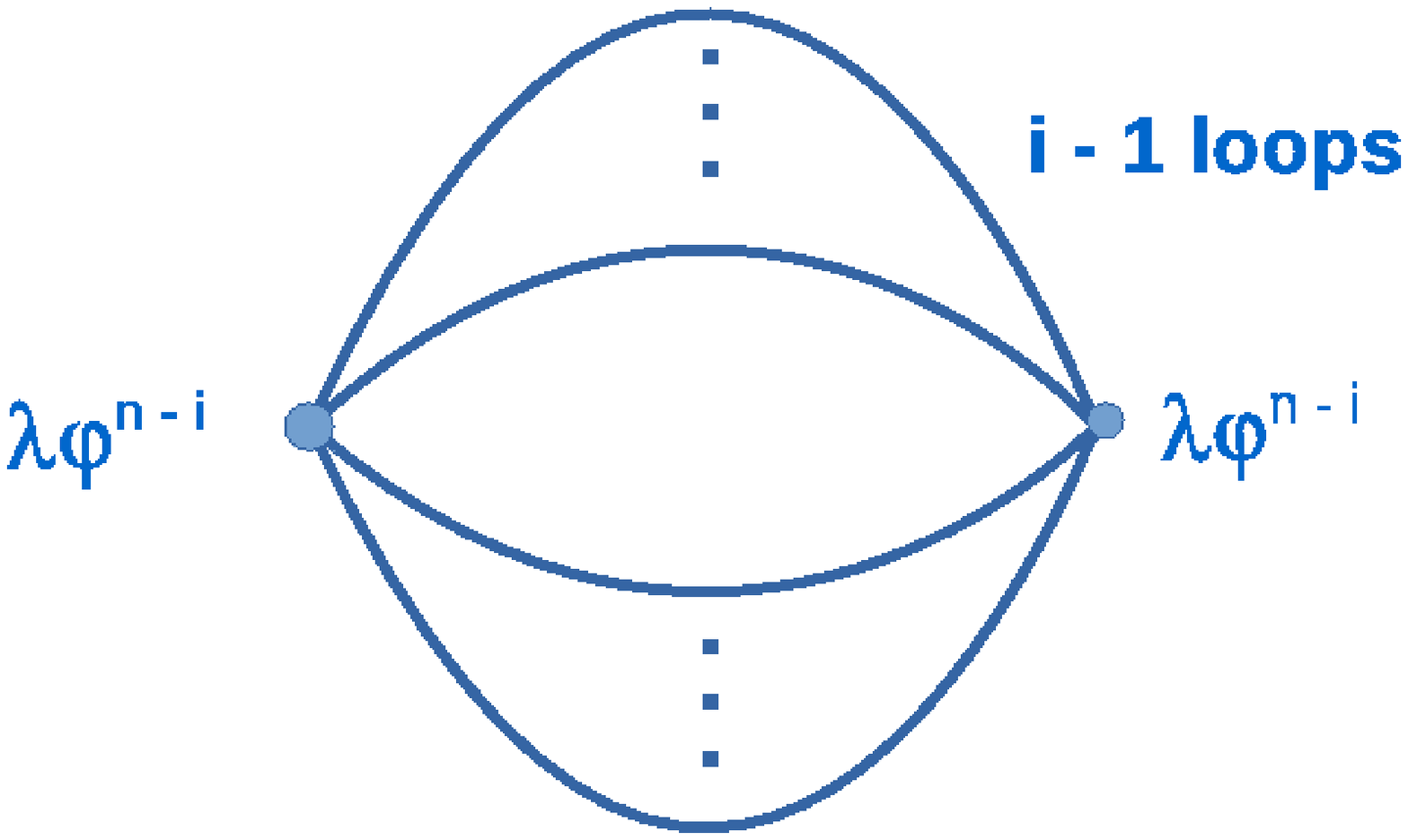} & $+$ & 
\includegraphics[width=3cm]{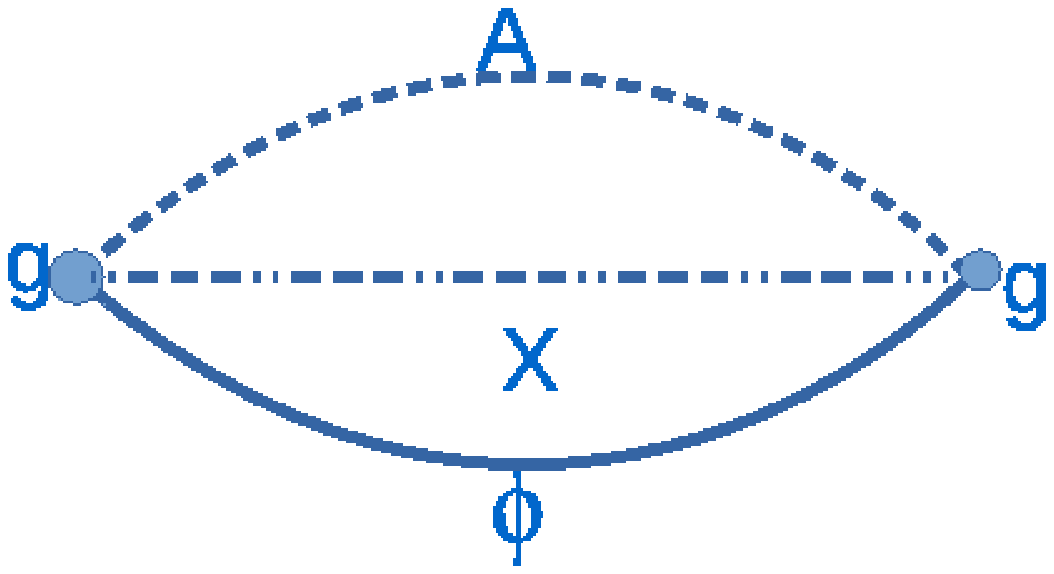} & $+ \ldots$ \\ & & & & & & \\
$N_1 = C_i^n i!!$ & & \multicolumn{2}{l}{$N_2 = i! (C_i^n)^2$} & & &  
\end{tabular}
}
\caption{Diagrams contributing to $\Gamma_2 (\varphi, G)$. \label{fig:gammaeffdiag}}
\end{figure}

\begin{figure}
{\color{darkblue}
\begin{tabular}{p{0.5cm}p{3cm}p{2cm}p{3cm}p{2cm}p{0.2cm}p{0.5cm}p{1.5cm}p{0.5cm}}
$\frac{\partial\Gamma_2}{\partial \varphi} = $ & $\sum\limits_{i=2}^{n-1} (n-i) N_1$ & 
\hspace{-0.5cm}\includegraphics[width=2cm]{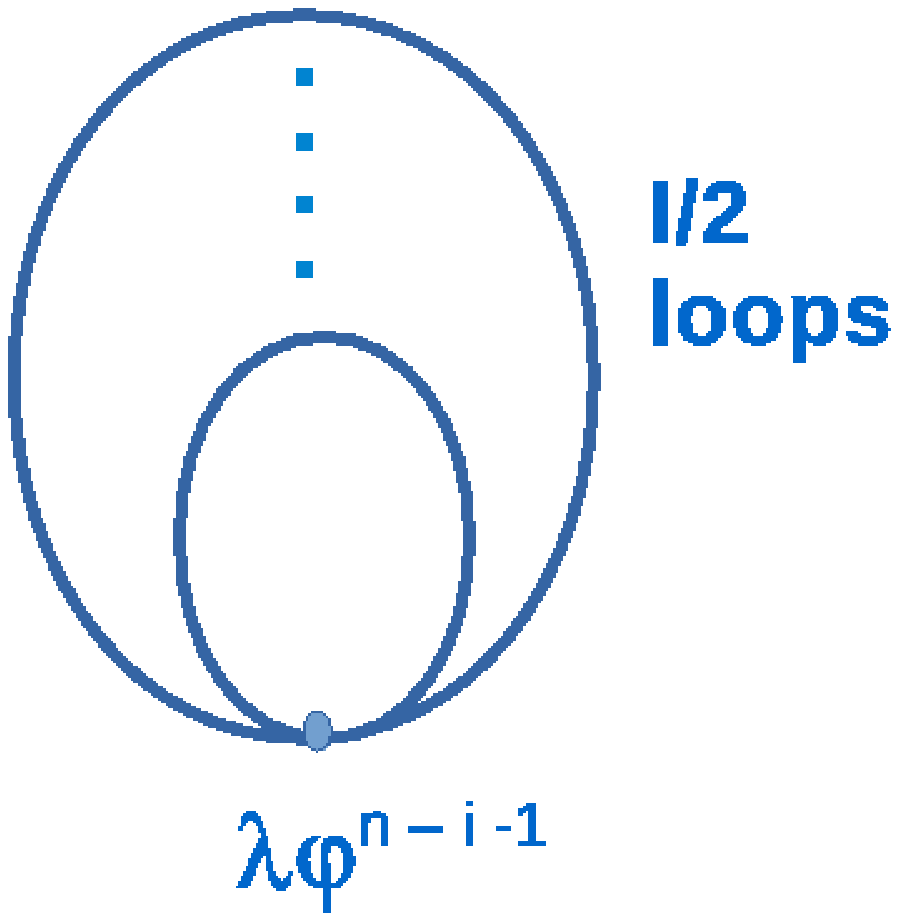} & \hspace{-0.5cm}
$+ \sum\limits_{i=2}^{n-1} (n-i) N_2$ & \includegraphics[width=2.5cm]{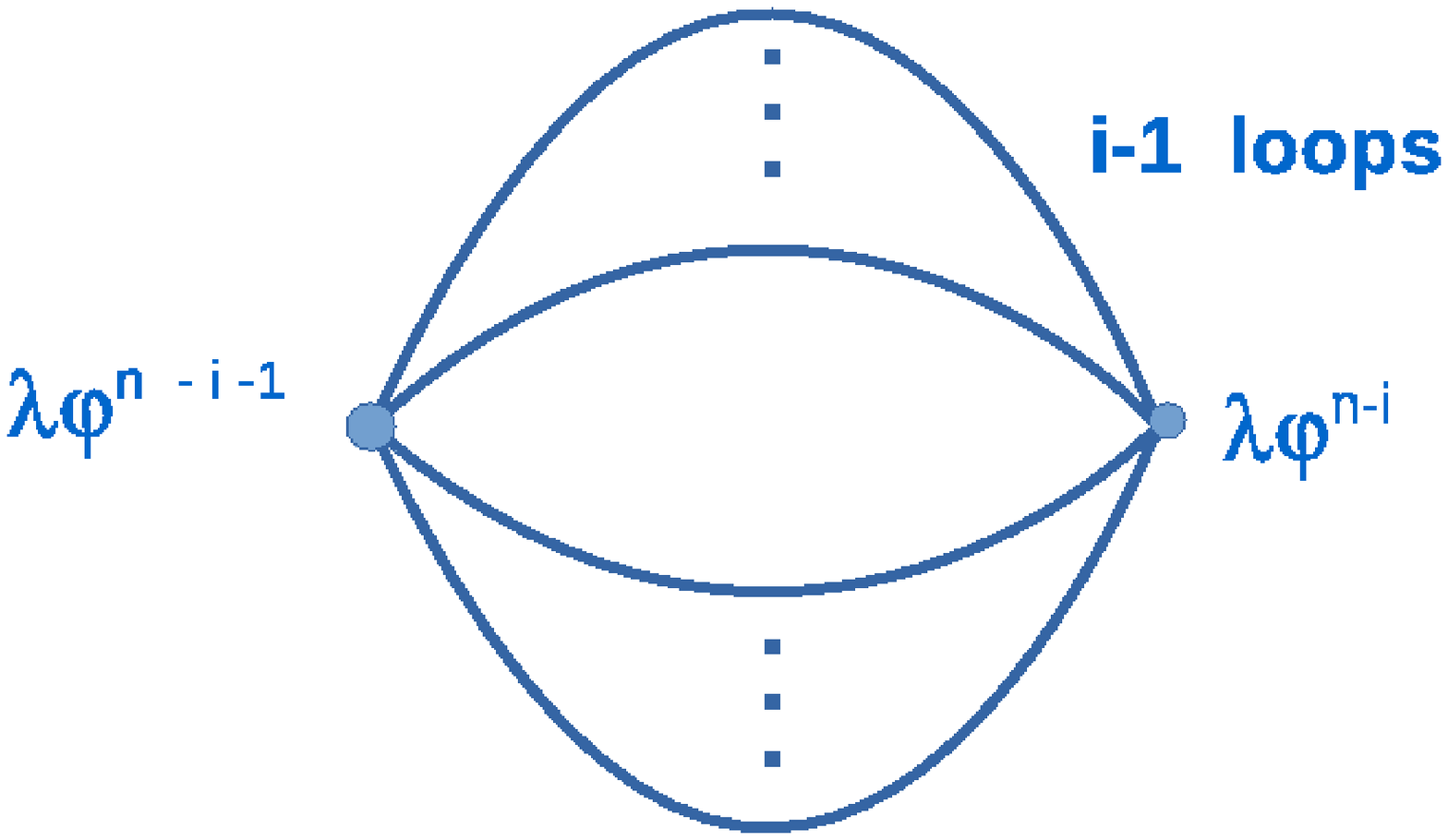} & $+$ &
 & \includegraphics[width=2cm]{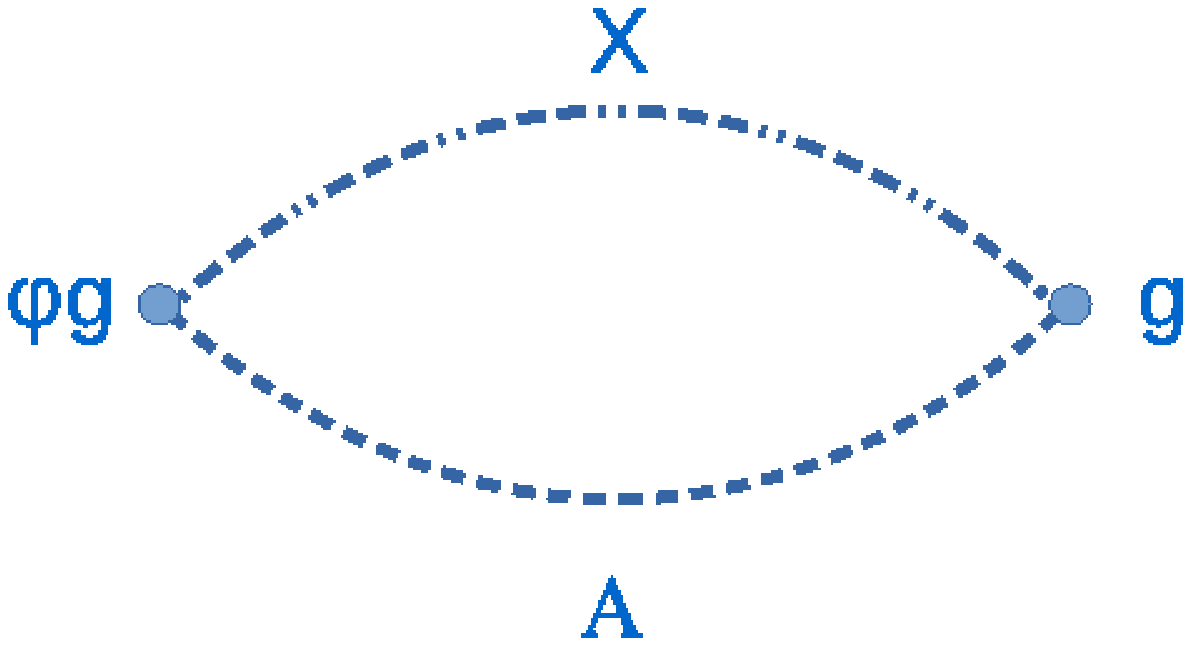} & $ + \ldots$
\end{tabular}
}
\caption{Diagrams contributing to $\partial \Gamma_2 (\varphi, G)/\partial \varphi$. Coefficients 
$N_1$ and $N_2$ are defined in Fig. \ref{fig:gammaeffdiag}. \label{fig:condensatediag}}
\end{figure}

Evolution equations of the condensate and propagators can be obtained from integration of 
effective action $\Gamma [\varphi,{G_{\alpha\beta}}],~\alpha,~\beta \in {X,~A,~\Phi}$ and are the 
followings:
\be
\frac{1}{\sqrt{-g}}{\partial}_{\mu}(\sqrt{-g} g^{\mu\nu}{\partial}_{\nu}
\varphi) + m_{\Phi}^2 \varphi + \frac{\lambda}{n!}\sum_{i=0}^{n-1} (i+1)
\binom{n}{i+1}{\varphi}^i\langle{\phi}^{n-i-1}\rangle - \mg \langle XA\rangle = 0 \label {dyneffa} \\ 
\ee
\bea
\frac{1}{\sqrt{-g}}{\partial}_{\mu}(\sqrt{-g} g^{\mu\nu}{\partial}_{\nu} + M_i^2(x)) ~ 
G_i^F(x,y) & = & -\int_\infty^{x^0}d^4z \sqrt {-g (z)} ~ \Pi_{ij}^\rho(x,z) ~ G_{ij}^F(z,y) + \nonumber \\
&& \int_\infty^{y^0}d^4z \sqrt {-g (z)} ~ \Pi_{ij}^F(x,z) ~ G_{ij}^\rho(z,y) \label{evolgf} \\
\frac{1}{\sqrt{-g}}{\partial}_{\mu}(\sqrt{-g} g^{\mu\nu}{\partial}_{\nu} + M_i^2(x)) ~ 
G_i^\rho(x,y) & = & -\int_{y^0}^{x^0}d^4z \sqrt {-g (z)} ~ \Pi_{ij}^\rho(x,z) ~ G_{ij}^\rho(z,y) 
\label{evolgrho}
\eea
\bea
&& M_\Phi^2 (x) = m_\Phi^2 + \frac{-i\lambda}{(n-2)!} \sum_{j=0}^{[n/2]-1} C^{n-2}_{2j} C^{2j}_2 
\varphi^{n-2j}(x) (G^F_{\Phi} (x,x))^j, \quad M_{X, A}^2 = m_{X, A}^2 \label {local2pi} \\
&& \Pi (\varphi,G) \equiv 2i \frac{\partial \Gamma_2 [\varphi,G]}{\partial G} \label{2piselfener}
\eea
where $\Pi (\varphi,G)$ is self-energy and $\Gamma_2$ is the contribution of 2PI diagrams in the 
effective action. Indices $i,j$ define the field species $i = X,~A,~\phi$ and $F$ and 
$\rho$ indicate symmetric and antisymmetric propagators, respectively. We should remind that above 
equations are exact at all perturbative order and truncation to a limited order arises when one 
calculates self-energies and effective masses up to a limited number of loops and vertices. 
In (\ref{local2pi}) $[n/2]$ means the integer part of $n/2$ and $C^i_j$ is the combinatory 
coefficient. These equations are written for an arbitrary metric $g_{\mu\nu}$ which is treated as 
a classical field and we use semi-classical Einstein equation~\cite{2picurve} for its evolution. 
Figure \ref{fig:gammaeffdiag} shows lowest order 2PI diagrams contributing to the effective action. 
Each diagram should be understood as a closed double path and propagators and vertices has indices 
indicating on which path they are determined. Here, for the sake of simplicity we omit path 
indices. Figures \ref{fig:gammaeffdiag} and \ref {fig:condensatediag} shows 2PI 
diagrams contributing in evolution of condensate and propagators, respectively.

\begin{figure}
{\color{darkblue}
\begin{tabular}{p{0.5cm}p{2cm}p{2cm}p{3cm}p{2cm}p{0.5cm}p{2cm}p{0.5cm}}
$\frac{\partial\Gamma_2}{\partial G_\phi} = $ & $\sum\limits_{i=2}^{n-1} \frac {iN_1}{2}$ & 
\includegraphics[width=2cm]{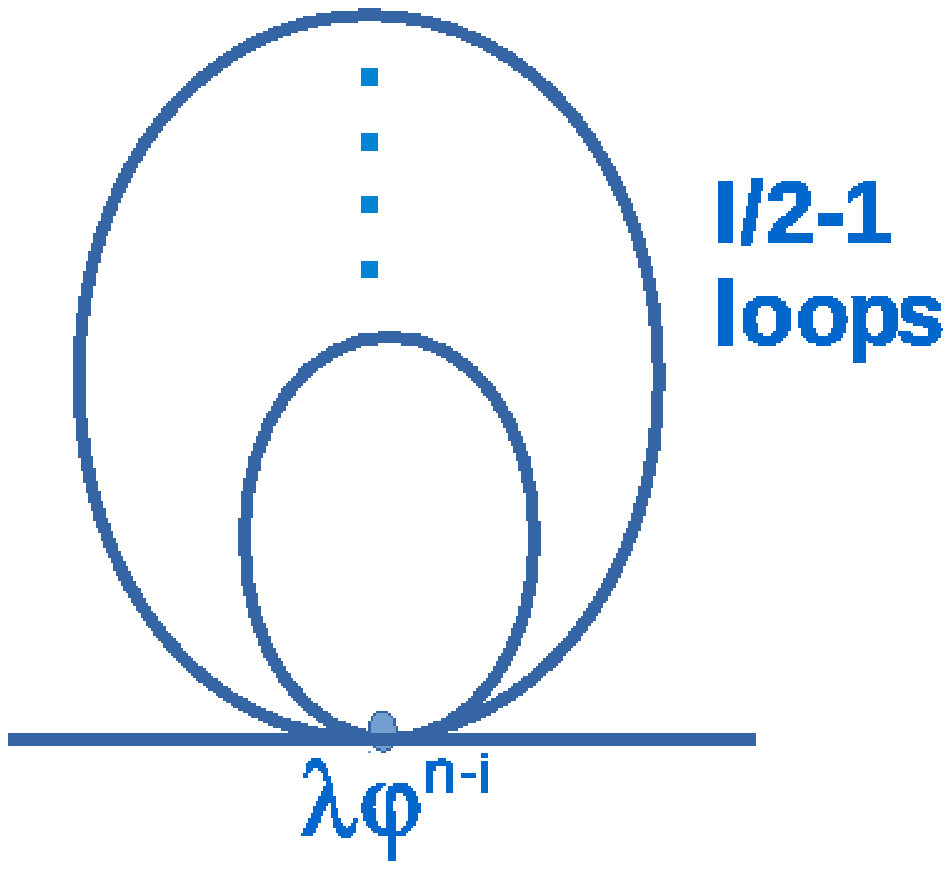} & $+ \sum\limits_{i=2}^{n-1} (i-1) N_2$ & \includegraphics[width=2cm]{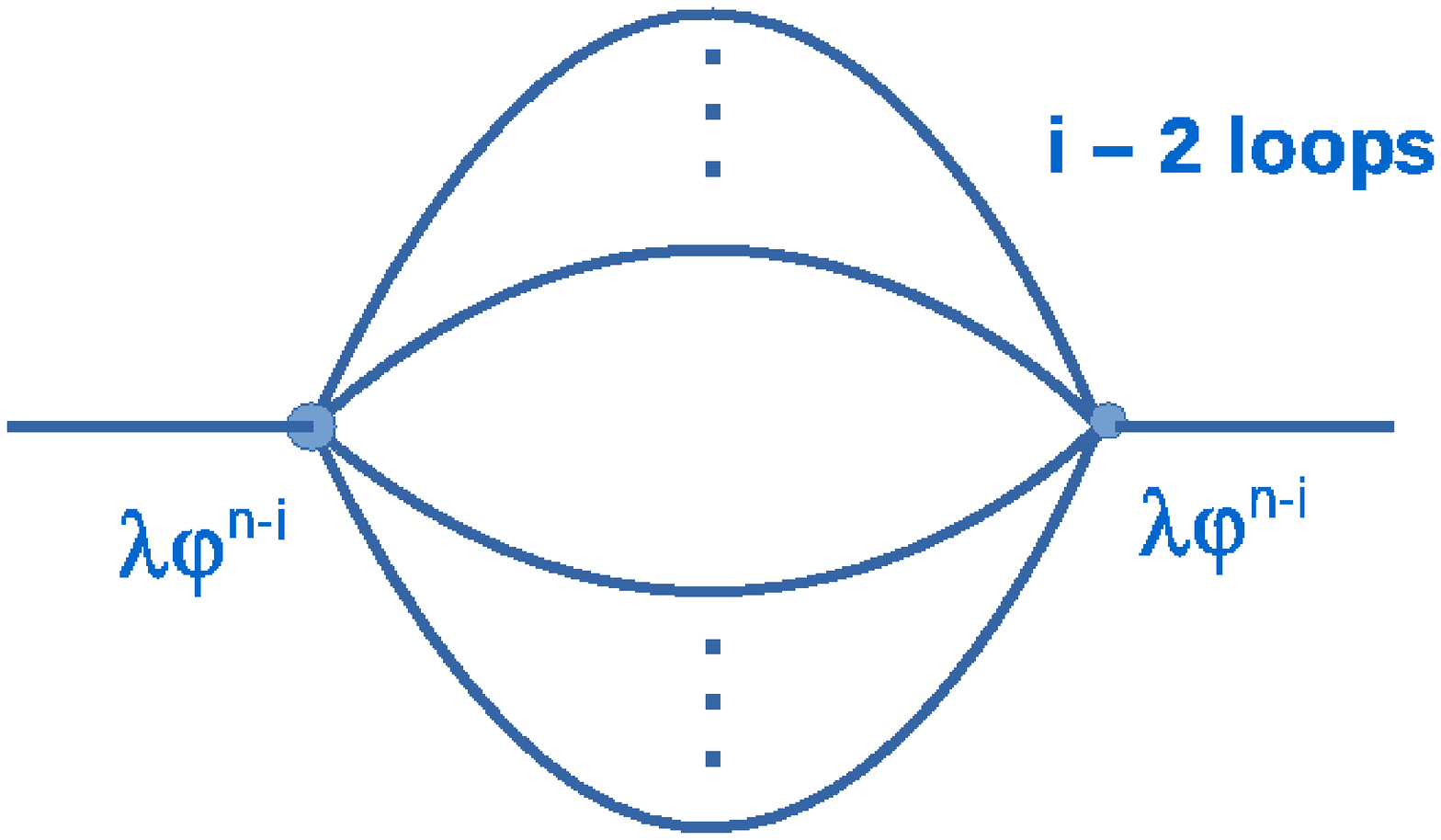} & $+$ & \includegraphics[width=2cm]{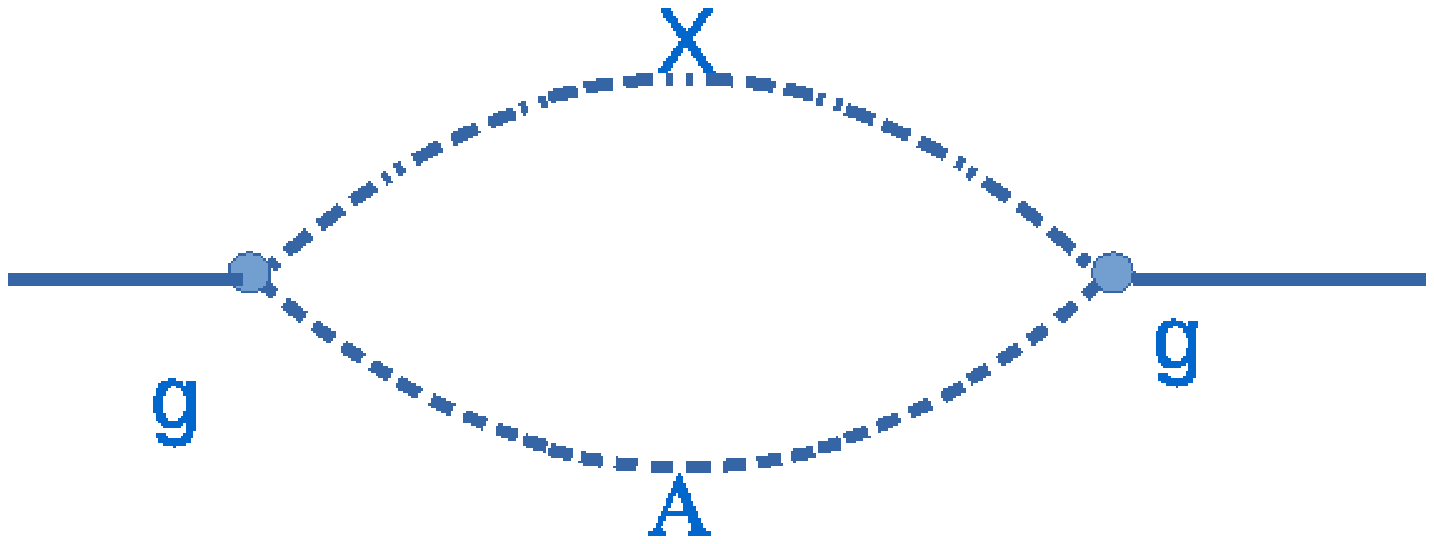} & $\quad + \ldots$ \\
$\frac{\partial\Gamma_2}{\partial G_X} = $ & \includegraphics[width=2cm]{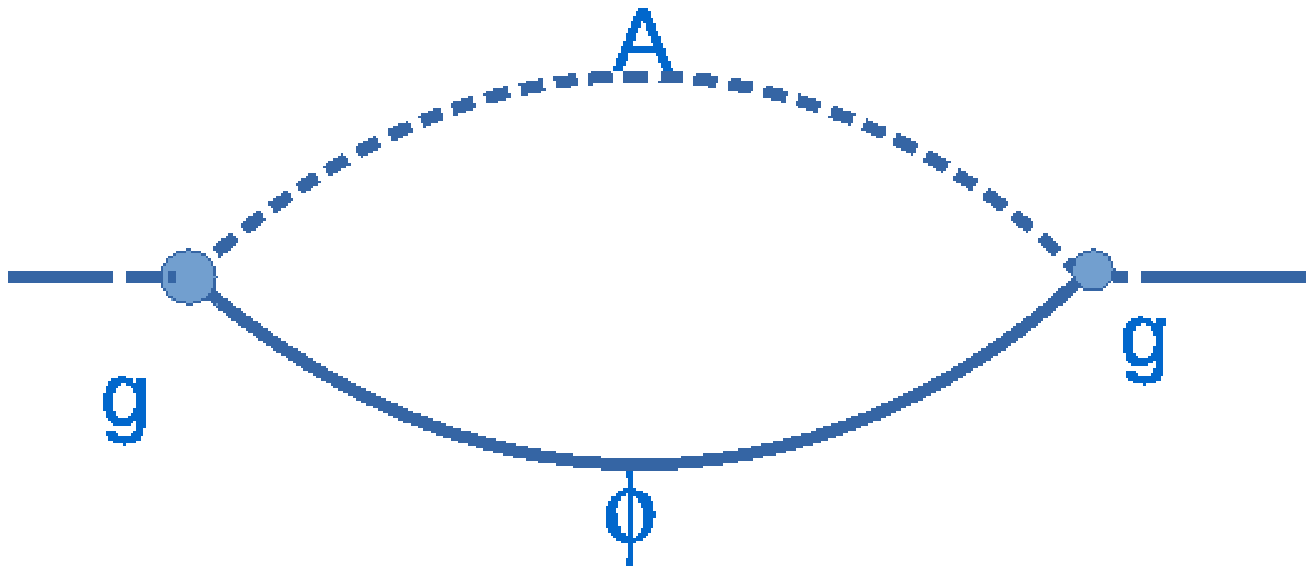} & $\frac{\partial\Gamma_2}{\partial G_A} = $ & \includegraphics[width=2cm]{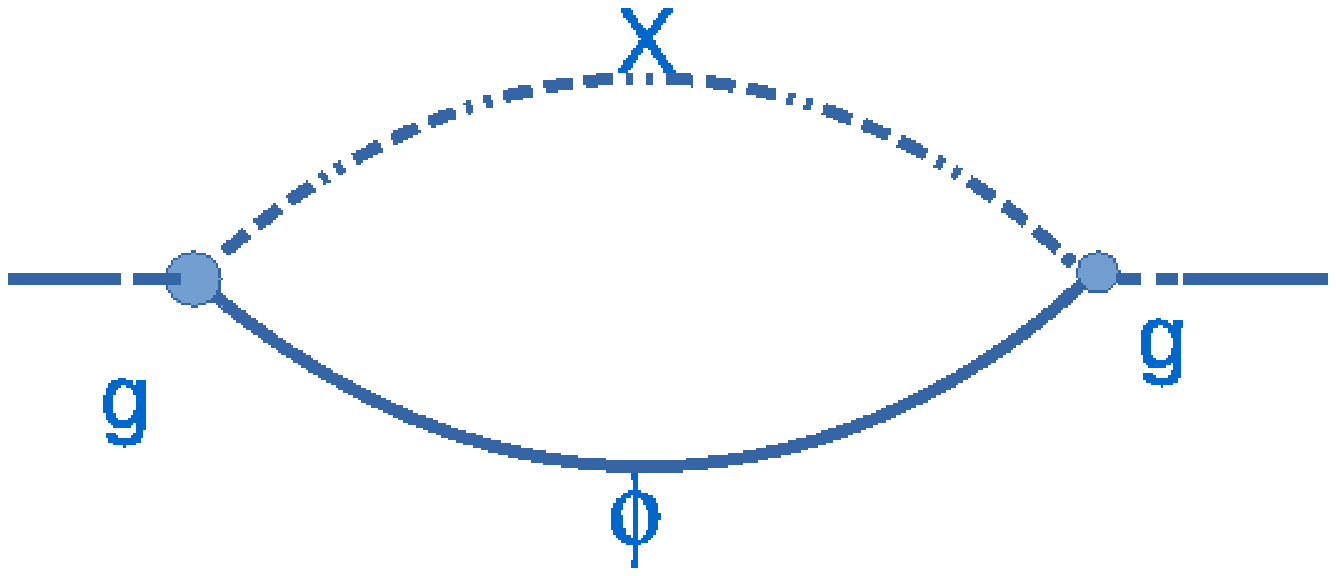}
\end{tabular}
}
\caption{Diagrams contributing to $\partial \Gamma_2 (\varphi, G)/\partial G_\phi$, $\partial 
\Gamma_2 (\varphi, G)/\partial G_X$, and $\partial \Gamma_2 (\varphi, G)/\partial G_A$. Coefficients 
$N_1$ and $N_2$ are defined in Fig. \ref{fig:gammaeffdiag}. The tadpole diagrams only contribute to 
the  mass term (\ref{local2pi}) and do not appear in the r.h.s. of equations (\ref{evolgf}) and 
(\ref{evolgrho}). \label{fig:propagdiag}}
\end{figure}

\subsection{Approximate analytical solution in radiation and matter domination eras}
Equations (\ref{dyneffa}-\ref{evolgrho}) cannot be solved analytically. The lowest order of 
correction is when free propagators are used, which corresponds to neglecting the 
right hand side of equations (\ref{evolgf}) and (\ref{evolgrho}), and considering only tree 
diagrams (which are not shown here) for determination of expectation values in the condensate 
evolution equation (\ref{dyneffa}). Moreover, rather than evolving the metric consistently, 
we assume that it is controlled by a component external to the model. The reason is the coupling 
of the expansion factor to fields, even for a homogeneous FLRW geometry. This makes analytical 
solution of evolution equations impossible.

With these simplifications we solve evolution equations (see~\cite{houricond} for details) with 
an initial condition at radiation domination era and zero amplitude for the condensate. We observe 
that the amplitude of the condensate grows exponentially in a similar manner to preheating 
resonance. Obviously, this exponential gross cannot continue for long time otherwise it over closes 
the Universe. However, under simplifications explained above it is impossible to take the 
backreaction into account. Notably, interaction between condensate and non-condensed component, 
that is quantum fluctuations behaving approximately as free particles, is crucial for backreaction 
process and control of unlimited gross of condensate.

Next we assume a matter domination era and evolve the expansion rate accordingly. In this epoch 
even the simplest evolution equations in which all interactions and quantum corrections are 
neglected can be solved analytically only for special cases of zero mass or zero momentum Fourier
modes. Using these special solutions and WKB approximation, it is shown~\cite{houricond} that when 
interactions are neglected or linearized the amplitude of spatial Fourier transform of the 
condensate decreases with time as $\varphi_k \propto t_0/t$ meaning that condensate does not 
survive acceleration. Indeed this evolutionary behaviour leads to an equation of state 
$w \lesssim 0.7$ for the condensate - presumably dark energy - which is already excluded by 
observations. Therefore, to have a better idea about the exact evolution of condensate one has to 
solve the full nonlinear evolution equation of condensate. Evidently this is impossible, but some 
simplification plus power counting show that the equation includes terms similar to an effective 
potential with negative power which is a necessary condition for existence of tracking 
solutions~\cite{trackingsol}. We find that for a self-interacting polynomial potential of order $n$, 
quantum corrections of order $i$ have negative exponents and close to constant coefficients if:
\be
\quad 17-6n+2i \geqslant 0, \quad i < n-1 
\ee
It is easy to verify that this condition is satisfied for $n \lesssim 4$, i.e. the only 
renormalizable scalar field models in 4D spacetimes. This conclusion which arises without 
renormalization of the model is by itself very interesting. However, giving large amount of 
simplifications performed for obtaining it, reduces the confidence, and therefore it needs 
confirmation with more robust analytical or numerical calculations.

\subsection{Numerical simulations}
\begin{figure}
\begin{center}
\includegraphics[width=5cm,angle=-90]{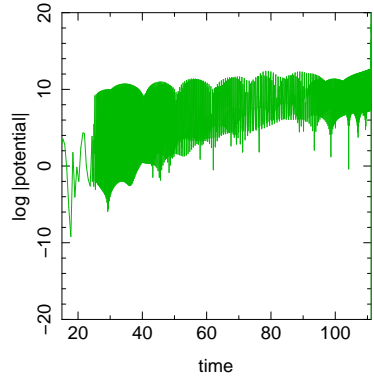}
\end{center}
\caption{Effective potential of a light scalar field. Initial conditions of the simulation is 
defined for pre-inflationary epoch with a null initial value for the amplitude of the condensate. 
\label{simulinfcond}}
\end{figure}

Numerical simulation of the model described in the previous subsections is a work in progress. 
Figure \ref{simulinfcond} shows an example of these simulations in which the initial condition is 
fixed for pre-inflationary Universe, with $\langle \phi (t_0) \rangle = 0$ and no free particle, 
i.e. a vacuum state. It shows that in contrast to the case of radiation domination universe, after 
an initial increase the effective potential of condensate arrives to a saturated state and 
produces an inflation. Ultimately simulations should be continued for enough long time to achieve 
an $N \sim 60$ e-folding before it decreases and transfer its energy to a free $\Phi$ particles 
and other species. For the time being, due to numerical errors our simulations cannot achieve this 
goal and we are working on its improvement. Notably, it is plausible that limited order of quantum 
corrections considered in the theoretical formulation is not enough for dealing with ultra soft 
modes. Indeed theoretical investigations show that infrared (IR) modes of massless and light fields 
in de Sitter space may have an IR singularity~\cite{desitterirsingular} when treated with 
perturbative techniques and need non-perturbative treatment. An example of an analogue situation 
is {\it small-x} regime in Deep Inelastic Scattering of hadrons and pair-production in strong 
electric or magnetic fields~\cite{pairprodmag}. A method analogue to Color Glass Condensate 
(CGC)~\cite{qcdcgc} used in QCD is developed for curved spacetimes~\cite{hourismallx} and may be 
applicable to cosmological IR problem. 

\section{Outline}
Coherent states are fully quantum states that behave collectively like a classical field. They may 
had a crucial role during inflation and reheating, and present expansion of the Universe may be 
also due to coherent superposition at cosmological distances. In this case, in one hand 
cosmological coherent states may be considered as ultimate environment for decoherence of 
quantum states. Moreover, if dark energy is a quantum condensate, its present dominant 
contribution to total energy of the Universe means that its constituents are mainly in a quantum 
superposition state rather than being decohered and classical. The verification of such a claim is 
not easy. Nonetheless, better understanding of early Universe physics through observations of 
the CMB anisotropies and properties of dark energy, as well as theoretical works, may allow to 
find a mean for verification of large scale quantum properties of the Universe.

\end{document}